\documentclass[aps,twocolumn,epsfig,graphics,showpacs,floatfix,superscriptaddress]{revtex4}
\usepackage{amsmath,amsfonts,amssymb,graphics,graphicx,epsfig,color,times,bbm}
\usepackage[latin1]{inputenc}

\graphicspath{{Bilder/}}

\newcommand{\ket}[1]{\left| #1 \right.\rangle}

\newcommand{\ad}{\hat{b}^\dagger}
\renewcommand{\a}{\hat{b}}
\newcommand{\nb}{\hat{n}}
\newcommand{\cd}{\hat{c}^\dagger}
\renewcommand{\c}{\hat{c}}
\newcommand{\nf}{\hat{m}}

\newcommand{\nQP}{\hat{Q}}

\usepackage[scanall]{psfrag}
\usepackage{subfigure}

\DeclareMathSymbol{\epsilon}{\mathord}{letters}{"22}
\DeclareMathSymbol{\theta}{\mathord}{letters}{"23}
\DeclareMathSymbol{\rho}{\mathord}{letters}{"25}
\DeclareMathSymbol{\phi}{\mathord}{letters}{"27}

\DeclareMathSymbol{\varepsilon}{\mathord}{letters}{"0F}
\DeclareMathSymbol{\vartheta}{\mathord}{letters}{"12}
\DeclareMathSymbol{\varphi}{\mathord}{letters}{"1E}
\DeclareMathSymbol{\varrho}{\mathord}{letters}{"1A}
\begin{document}

\newcommand{\nn}{{\mathbbm{N}}}
\newcommand{\rr}{{\mathbbm{R}}}
\newcommand{\cc}{{\mathbbm{C}}}
\newcommand{\id}{{\sf 1 \hspace{-0.3ex} \rule{0.1ex}{1.52ex}\rule[-.01ex]{0.3ex}{0.1ex}}}
\newcommand{\me}{\mathrm{e}}
\newcommand{\mi}{\mathrm{i}}
\newcommand{\md}{\mathrm{d}}
\renewcommand{\vec}[1]{\text{\boldmath$#1$}}

\title{The one-dimensional Bose-Fermi-Hubbard model in the heavy-fermion limit}

\author{A.\ Mering and M.\ Fleischhauer}
\affiliation{Fachbereich Physik, Technische Universit\"at Kaiserslautern, D-67663 Kaiserslautern, Germany}






\begin{abstract}
We study the phase diagram of the zero-temperature, one-dimensional Bose-Fermi-Hubbard model for fixed fermion density in the limit of small fermionic hopping. This model can be regarded as an instance of a disordered Bose-Hubbard model with dichotomic values of the stochastic variables. Phase boundaries between compressible, incompressible (Mott-insulating) and partially compressible phases are derived analytically  within a generalized strong-coupling expansion and numerically using density matrix renormalization group (DMRG) methods. We show that  first-order correlations in the partially compressible phases decay exponentially, indicating a glass-type behaviour. Fluctuations within the respective incompressible phases are determined using perturbation theory and are compared to DMRG results.
\end{abstract}


\date{\today}

\maketitle

\section{Introduction}

Ultracold atoms in optical lattices provide an experimentally 
accessible toolbox for simulating strongly correlated quantum systems
\cite{lit:Bloch-RMP-2007,lit:Jaksch-AnnPhys-2005,lit:Jaksch-PRL-1998,lit:Fisher-PRB-1988,lit:Greiner-Nature-2002}. 
The interaction of the atoms gives rise to local Hamiltonians on a lattice 
that can be characterized in their microscopic details. Moreover,
by means of mixtures different species, Feshbach resonances, or additional
optical lattices, an unprecedented control over system parameters
can be achieved. Following first experiments showing a Mott-superfluid
phase transition \cite{lit:Greiner-Nature-2002} in a bosonic system \cite{lit:Fisher-PRB-1988,lit:Jaksch-PRL-1998}, 
a plethora of systems of cold atoms 
have been studied. This includes mixtures of bosonic and fermionic atoms,
 -- studied in experiments \cite{lit:Ketterle-PRL-2002,lit:Inguscio-PRL-2004,lit:Esslinger-PRL-2006,lit:Ospelkaus-PRL-2006}
 and theory \cite{lit:Albus-PRA-2003,lit:Lewenstein-PRL-2004,lit:Lewenstein-OptComm-2004,lit:Cramer-PRL-2004,lit:Pazy-PRA-2005,lit:Mathey-PRL-2004,lit:Demler-PRA-2006,lit:Roth-PRA-2004,lit:Buechler-PRL-2003,lit:Buechler-PRA-2004,lit:Pollet-PRL-2006,lit:Pollet-condmat-2006} -- 
giving rise to a rich phase diagram and complex physics, including 
fermion pairing, phase separation, density waves, and supersolids.

Early theoretical studies of the BFHM within mean-field and Gutzwiller decoupling approaches
\cite{lit:Albus-PRA-2003,lit:Lewenstein-OptComm-2004,lit:Cramer-PRL-2004} as well as exact numerical
diagonalization \cite{lit:Roth-PRA-2004}
revealed the existence of Mott-insulating  (incompressible) phases with incommensurate boson
filling.  In comparison to the Bose-Hubbard model where the incompressible phases are entirely characterized by
the local boson number, the corresponding phases for Bose-Fermi mixtures display a much richer 
internal structure. A rather complete description of these 
phases can be obtained using a composite-fermion picture \cite{lit:Lewenstein-PRL-2004}, which 
predicts density waves with integer filling, the formation of composite-fermion domains (phase separation),
composite Fermi liquids and BCS-type pairing. The properties of 1D Bose-Fermi mixtures in the 
compressible phases were analyzed in terms of fermionic polarons using a 
bosonization approach  \cite{lit:Mathey-PRL-2004}. In two spatial dimensions the existence
of super-solid phases  was predicted  \cite{lit:Buechler-PRL-2003} which is characterized by the 
simultaneous presence of a density wave and long-range off-diagonal order for the bosons. 
The persistence of a density wave with noninteger fillings in the compressible phases was
shown in \cite{lit:Pazy-PRA-2005}. There are also a few exact numerical studies using both
quantum Mote-Carlo  \cite{lit:Pollet-PRL-2006} and DMRG calculations \cite{lit:Pollet-condmat-2006}.

In the present work, we consider a Bose-Fermi mixture in a one-dimensional, deep periodic lattice described by the
Bose-Fermi-Hubbard model (BFHM). In particular we study the case of small fermionic hopping, where 
the presence or absence of a fermion at a lattice site results in a dichotomic random alteration of the
local potential for the bosons. We show that for this limiting case a rather accurate prediction
of the incompressible (Mott-insulating) phases is possible using a generalized strong-coupling approach.
To verify this approach we perform numerical simulations using the density-matrix renormalization group (DMRG)
\cite{lit:Schollwoeck-RMP-2005}. 
We predict the existence of partially compressible phases and provide numerical evidence that they have a Bose-glass
character. Finally we calculate local properties in the incompressible phases and draw conclusions
about the validity of effective theories. 

\section{The model}\label{model}

We consider a mixture of ultra-cold, spin polarized fermions and bosons in an
optical lattice. In the tight binding limit of a deep lattice potential, 
the
system can be described by the {\it Bose-Fermi Hubbard model} 
\cite{lit:Albus-PRA-2003}. 
We here consider a semi-canonical model, in which the number of fermions is hold constant, 
but in which we allow
for fluctuations of the total number of bosons  
determined by a chemical potential $\mu$. The corresponding 
Hamiltonian reads 
\begin{eqnarray}\label{eq:BFHM}
	\hat{H}&=&-J_B\sum_j\left(\ad_j\a_{j+1}+\ad_{j+1}\a_{j}\right)-\mu\sum_j \nb_j\\
	&-&J_F\sum_j\left(\cd_j\c_{j+1}+ \cd_{j+1}\c_{j}\right)\nonumber\\
	&+&\frac{U}{2}\sum_j\nb_j\left(\nb_j-1\right)+V\sum_j\nb_j\nf_j.\nonumber
\end{eqnarray}
Here, $\c_j$ and $\a_j$ are the annihilation operators of the fermions and bosons at lattice site
$j$, respectively, and $\nb_j=\ad_j\a_j$, 
$\nf_j=\cd_j\c_j$ the corresponding number operators. 
The particles can tunnel from one lattice site
to a neighboring one, the rate of which is described by $J_B$  and $J_F$ 
for bosons and fermions, respectively. $V$ 
is the on-site interaction strength between the two species, while $U$ accounts for 
intra-species repulsion of bosons, which will define our energy scale and we 
set henceforth $U=1$. 

Throughout the present work, we will 
focus on the case of heavy, immobile fermions, i.e., 
we consider the limit in which $J_F =0$ is a good approximation. 
In this case the effect of the fermions 
reduces to a dichotomic random potential at site $j$ 
for the bosons, depending on whether 
a fermion is at site $j$ or not. This means that the local potential is altered
by
\begin{equation}
	\delta\mu_j =\left\{
	\begin{array}{ll}
	V, & \text{ if a fermion is present at site $j$,} \\
	0, & \text{ otherwise}.
	\end{array}\right.
\end{equation}
We will systematically
investigate to what extent this limit of the Bose-Fermi-Hubbard model can be described as an
specific instance of a 
{\it disordered Bose-Hubbard} model, 
\begin{eqnarray}
	\nonumber
	\hat{H}&=&-J_B\sum_j\left(\ad_j\a_{j+1}+\ad_{j+1}\a_{j}\right)-\sum_j\Bigl(\mu-\delta\mu_j\Bigr) \nb_j\\
	&+&\frac{1}{2}\sum_j\nb_j\left(\nb_j-1\right)	\label{eq:BFHM-random}
\end{eqnarray}
We will see that this simple model shows on the one hand
important features of the full  Bose-Fermi-Hubbard model. On the other hand, 
we will see that this leads to important qualitative differences to the phase diagram
of the disordered Bose-Hubbard model with continuously distributed on-site disorder, 
as studied in Refs.\ \cite{lit:Fisher-PRB-1988,lit:Scholl-EPL-1999}. 
Depending on the physical situation of interest we will consider two cases
of disorder: If the fermionic tunneling is small but sufficiently large such that
on the time scales of interest relaxation to the state of total minimum energy 
is possible, the fermion induced
disorder is referred to as being {\it annealed}. In this 
case the ground state is determined by
minimization over all possible fermion distributions. If the fermion 
tunneling is too
slow or the temperature too high the disorder is an actually random distribution
called {\it quenched}.

\section{Compressible and incompressible phases}
\label{phases_constant_fermi_filling}

In this section we derive the phase diagram of the BFHM with immobile fermions. 
More specifically, we will approximate the boundaries between compressible and 
incompressible phases employing a generalization of  
the familiar strong-coupling expansion 
\cite{lit:Freericks-PRB-1996} to the present case of bosons with a modified potential
due to the presence of fermions. This will be compared to
the predictions of several instances of mean-field approaches 
\cite{lit:Cramer-PRL-2004,lit:Lewenstein-OptComm-2004}. 
Furthermore, a comparison with numerical results in one spatial dimension obtained by a 
DMRG computation will be given. Our strong-coupling expansion reveals the 
existence of novel phases whose character will be discussed in the 
subsequent section. 

\subsection{Ultra-deep lattices}

We first discuss the simple 
case of an ultra-deep lattice for the bosons, 
such that their hopping can be
neglected. 
In this situation where  $J_F=J_B = 0$ 
is a good approximation, the Hamiltonian becomes diagonal in the
occupation number basis. This basis will be denoted as  
$\{|n_1, \cdots, n_N\rangle |  m_1, \cdots, m_N\rangle \}$, where $m_j=0,1$
denotes the number of fermions at site $j$ and $n_j=0,1,...$ the 
corresponding number of bosons. 
$N$ labels the total number of lattice sites in the one-dimensional system.
The problem of finding the ground state reduces to 
identifying product states with the lowest energy. 
By fixing the total number of fermions $N_F=N\rho_F$,
this amounts to minimizing  
\begin{eqnarray}
	E &= &
	 \frac{1}{2}\sum_{j} n_j (n_j-1)- 
	(\mu-V)
	\sum_{j\in {\cal F}} n_j - 
	 \mu 
	 \sum_{j \in {\cal N}}  n_j. \nonumber
\end{eqnarray}
where ${\cal F}$ denotes the set of $\rho_F N = N_F$ sites with a fermion and
${\cal N}$ the set of $(1-\rho_F) N = N- N_F$ sites without a fermion, $\rho_F$ denoting the 
fermionic filling factor.
The energy is obviously degenerate for all fermion distributions 
and the ground state is given by an equal mixture of all states with 
state vectors
\begin{equation}
 |\psi_0\rangle = \bigotimes_{i\in {\cal F}} |{n}_1,1\rangle \bigotimes_{j\in{\cal N}} |{n}_0,0\rangle.
\end{equation}
Here,
\begin{equation}
	\label{occ_number}
	{n}_1=\max\Bigl\{0,[1/2+(\mu-V)]\Bigr\},\,
	{n}_0=\max\Bigl\{0,[1/2+\mu]\Bigr\},
\end{equation}
is the local boson number for sites with ({$\cal F$}) or without ({${\cal N}$}) a fermion and
 $[.]$ denotes the closest integer bracket. In other words, the degenerate 
states with lowest energy will have
$\rho_FN$ sites with ${n}_1$ bosons and one fermion and
$N(1-\rho_F)$ sites with ${n}_0$ bosons and no fermion. 
For the case of zero or unity fermion filling, $\rho_F=1$ the situation becomes 
particularly simple as we encounter the pure Bose-Hubbard model with an 
effective chemical potential $\mu^{\rm eff}=\mu-V\rho_F$.

\begin{figure}[thb]
\epsfig{file=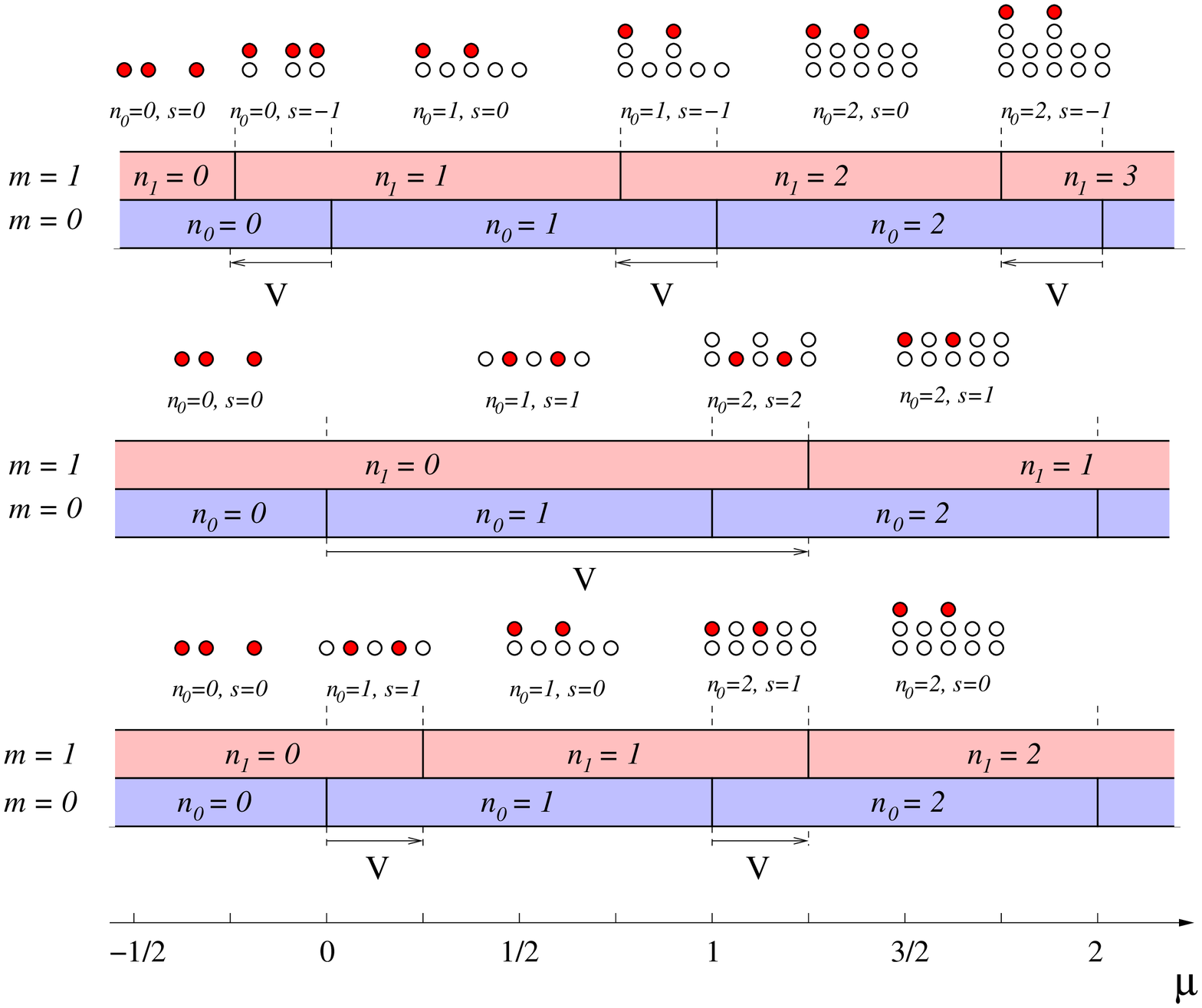,width=8 cm}
\caption{(Color online) Phases of BFHM for $J_B=J_F=0$ for different inter-species couplings
$0<V<1$ (lowest diagram), $1<V<2$ (middle diagram), $-1<V<0$ (upper diagram), and $U=1$. $n$ indicates the number of bosons (empty circles) 
at the site, $m$ the number of fermions (red filled circles).
The horizontal red bars illustrate the boson number $n_1$ for sites with a fermion ($m=1$) as function
of the chemical potential, the horizontal blue bars correspondingly the boson number $n_0$ for sites
without a fermion ($m=0$), which is identical to the BHM.
The values of $\mu$ where a transition between different boson numbers $n_0$ occurs either at sites without
a fermion ($m=0$) or with a fermion ($m=1$) are quantum critical points.}
\label{fig:JB=0-Phase-diagram}
\end{figure}

Since ${n}_1$ and ${n}_0$ are integers there are adjacent intervals of $\mu$ where the 
occupation numbers do not change.  In these intervals the system is incompressible,
i.e., 
\begin{equation}
\frac{\partial \langle \sum_j {\hat n}_j\rangle }{\partial \mu} =0
\end{equation}
and the points between two intervals
are quantum critical points.
This behavior, illustrated in Fig.\ \ref{fig:JB=0-Phase-diagram}, is very similar to that
of the Bose-Hubbard model except that here the 
bosons can be incompressible even for non-integer filling $\rho_B$ as we have 
$\rho_B={n}_0+\rho_F({n}_1-{n}_0)$. 
Following Ref.\ \cite{lit:Lewenstein-PRL-2004} we label the difference 
${n}_0-{n}_1$ in the bosonic number mediated through the presence of a fermion by $s$.
The local ground state can either consist of ${n}_0$ bosons and no fermion or $n_1={n}_0-s$ 
bosons and one fermion. These state vectors will be denoted as 
$\ket{{n}_0,0}=\ket{0}$ and $\ket{{n}_0-s,1}=\ket{1}$. 
The value  of $s$ depends 
on $\mu$ and $V$ and can be a positive or negative integer. 
Both these vectors are eigenvectors
of the number operator 
\begin{equation}
	\nQP_j = \nb_j+s\nf_j
\end{equation}	
with the same integer eigenvalue $ n_0$ and 
$\langle \Delta \hat Q^2_j\rangle =0$.
Thus incompressible phases have a commensurate number $\nQP$
and can be characterized by the two integers  $n_0$ and $s$ 
Since ${n}_0$ and ${n}_1$ 
are integers and increase monotonically with $\mu$ , there is a jump in the total number
of bosons when moving from one incompressible to the adjacent one.
 All systems with boson number in between these values are critical
and have the same chemical potential since $J_B=0$.
The average boson number per site in the incompressible phases does not have to be
an integer, however. The existence of Mott phases with non-commensurate boson number is
a direct consequence of the dichotomic character of the fermion induced disorder.
A similar behavior has been predicted for superlattices, which can be considered
as dichotomic disorder in the special case of anti-clustering \cite{lit:Roth-PRA-2003,lit:Buonsante-PRA-2004}.
In general Mott-insulating phases with incommensurate boson numbers exist for
any disorder distribution that is non-continuous.

\subsection{Minimum energy distribution of fermions for small bosonic hopping}
\label{fermion-minimization}

In order to understand the physics for disorder due to the presence of fermions we need to discuss the
influence of the distribution of fermions to the ground state energy.
The energetic degeneracy of different fermion distributions in the incompressible phases 
is lifted if a small bosonic hopping $J_B$ is taken into account. Near the quantum critical points
the boson hopping leads to the formation of possibly critical phases with growing extent.
We first restrict ourselves to regions where incompressibility
is maintained, i.e. sufficiently far away from the critical points. 

\begin{figure}[t]
\epsfig{file=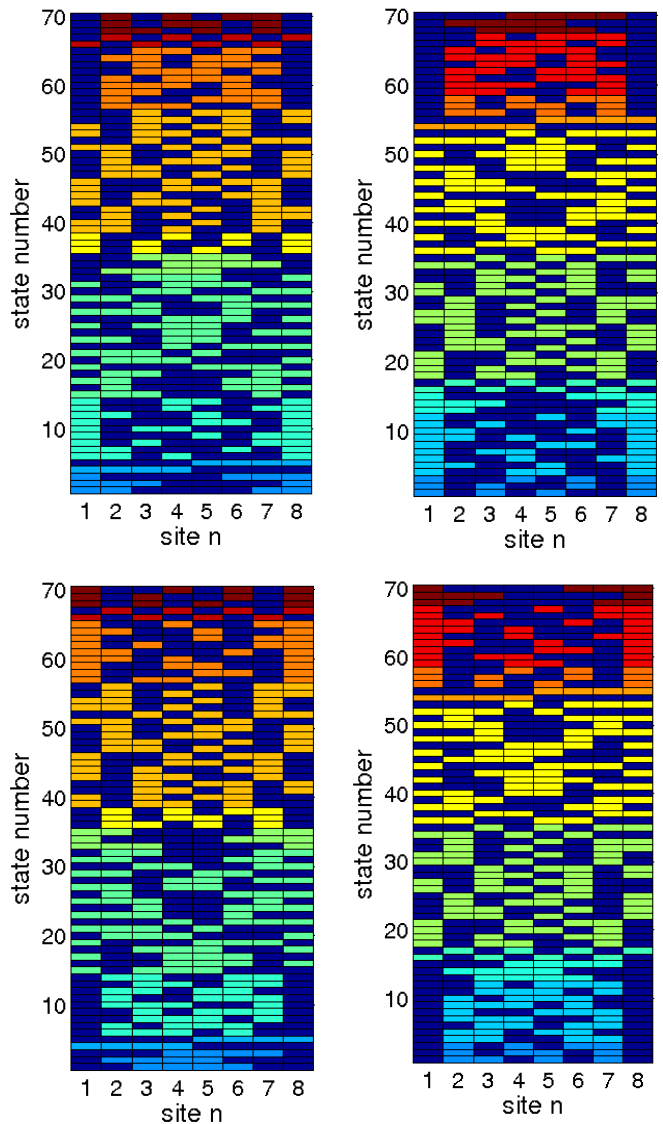,width=9cm}
\caption{(Color online) Fermion distributions ordered by ground state energy. Blue - lowest energy,
red -highest energy for $J_F=0$, $J_B=0.02$, $U=1$. \textit{top:}
attractive boundary, (i) $V=1.5, {n}_0=1, s=1$
i.e. $K_{\rm eff}=-0.002$, and (ii) $V=1.5, {n}_0=2, s=1$
i.e. $K_{\rm eff}=0.001$. \textit{bottom:} repulsive boundary (iii) $V=-1.5, {n}_0=0, s=-1$
i.e. $K_{\rm eff}=-0.002$, (iv) $V=-1.5, {n}_0=1, s=-1$
i.e. $K_{\rm eff}=0.001$.}
\label{fig:fermi-distr}
\end{figure}

In order to obtain a qualitative understanding of the effects of a finite bosonic hopping
we have performed a numerical perturbation calculation on a small lattice. 
Fig.\ref{fig:fermi-distr} shows
different distributions of 4 fermions over a lattice of 8 sites ordered according to their energy
for different parameters in 6th order perturbation.

One notices that the lowest energy states are either given by fermion distributions
with maximum mutual distance (anti-clustered configuration) or minimum
mutual distance (clustered configuration) modified by boundary effects. 
This behavior can in part be explained by the composite fermion picture
 introduced in \cite{lit:Lewenstein-PRL-2004}.
The composite fermions are defined for the phase $(n_0,s)$ by the annihilation operators:
\begin{eqnarray}
\hat f_i &=& \sqrt{\frac{(n_0-s)!}{n_0!}}\, \Bigl(\hat b_i^\dagger\Bigr)^s\, \hat c_i,\qquad\textrm{for}\enspace s\ge 0,\\
\hat f_i &=& \sqrt{\frac{n_0!}{(n_0-s)!}}\, \Bigl(\hat b_i\Bigr)^{-s}\, \hat c_i,\quad\enspace\textrm{for}\enspace s < 0.
\end{eqnarray}
For each $n_0$ and $s=1$, the full BFH-Hamiltonian, Eq.\ (\ref{eq:BFHM}), with $J_F=0$ gives
in second order in $J_B$ rise to the effective Hamiltonian \cite{lit:Lewenstein-PRL-2004}
\begin{equation}	
	\hat H_{\rm eff} = K_{\rm eff}\sum_{\langle j,k\rangle} (\hat f_j^\dagger\hat f_j)
	(\hat f_k^\dagger\hat f_k)\label{eq:composite-Ham},
\end{equation}
where $\langle.,.\rangle $ denotes nearest neighbors. Here, as  $J_F=0$, we find the 
effective coupling (note that again, $U=1$)
\begin{eqnarray}
	K_{\rm eff} &=&  4 J_B^2\Bigl[\frac{n_0(n_0+1-s)}{1-s+V}+\frac{(n_0-s)(n_0+1)}{1+s-V}
	 \nonumber \\ 
	&& \enspace -n_0(n_0+1)-(n_0-s)(n_0+1-s)\Bigr].
\end{eqnarray}
Composite fermions cannot occupy the same lattice site, 
but there may be nearest neighbor attraction ($K_{\rm eff}<0$) or
repulsion ($K_{\rm eff}>0$).  Associating a site with a composite fermion with a spin-up state
and a site without a fermion with spin down, Eq.\ (\ref{eq:composite-Ham}) corresponds to
the classical Ising model with fixed magnetization and anti-ferromagnetic ($K_{\rm eff}>0$) or ferromagnetic
coupling ($K_{\rm eff}<0$). 

As a consequence, to this order in perturbation theory, if $K_{\rm eff}<0$, the energy 
is smallest for fermion distributions that minimize the surface area of sites
with and without a fermion (referred to as {\it clustering}). In this setting, we can 
take the fermion distribution to form a block of occupied sites. 

The other regime is the one for $K_{\rm eff}>0$. Then, the fermions repel each other,
and they form a pattern with maximum number of boundaries for small $J_B$, referred
to as {\it anti-clustering}. That the fermions attain a distribution with maximum
distance cannot be explained by the effective model due to its perturbative nature.
In all of our numerical simulations using the density matrix renormalization
group (DMRG) we found however that a positive $K_{\rm eff}$ always lead to anti-clustering with maximum
distance. 

The ground state energies of the various fermionic distribution differ only by a small amount
which is on the order of $J_B^2/U$ or even higher powers.  Also for 
temperatures which are still small enough to treat the bosonic system with given disorder as an 
effective $T=0$ problem, but larger 
than the energy gap between different fermion distributions, i.e. for $J_B (J_B/U)^n \ll k_B T\ll J_B$, the various fermion distributions
will be equally populated. Thus it seems more natural to consider the case of quenched, random disorder rather than that of
annealed disorder.

\subsection{Compressible and incompressible phases for finite $J_B$}
\label{finite-JB}

We now discuss the boundaries of the incompressible phases for finite
bosonic hopping. To this end we extend the strong coupling expansion of Ref. 
\cite{lit:Freericks-PRB-1996} and complement the results with numerical DMRG simulations.
The strong coupling expansion provides a rather accurate
description for the Bose-Hubbard model even on a quantitative level. 

Let us consider a phase with $(n_0,s)$ and $N_F = \rho_F N$ fermions, i.e., a phase with
$N_F$ sites containing $n_0-s$ bosons and a fermion and $N-N_F$ sites with $n_0$ bosons.
The ground state vector for $J_B=0$ is then found to be
\begin{equation}
	\bigl|\psi_0\bigr\rangle = \bigotimes_{j\in{\cal F}} 
	\frac{{\hat c}_j^\dagger\bigl({\hat a}_j^\dagger\bigr)^{(n_0-s)}}{\sqrt{(n_0-	s)!}}
	\bigotimes_{j\in {\cal N}} \frac{\bigl({\hat a}_j^\dagger\bigr)^{n_0}}{\sqrt{n_0!}}\, 
	|0,\dots,0\rangle|0,\dots, 0\rangle .
\end{equation}
The energy density is given by
\begin{eqnarray}
	\epsilon_0 &=& \frac{U}{2} \Bigl[(1-\rho_F) n_0(n_0-1)+ \rho_F (n_0-s)(n_0-s-1)\Bigr]\nonumber\\
	& +& V \rho_F (n_0-s).
\end{eqnarray}
We now consider states with a single additional boson (bosonic hole). 
In contrast to the actual Bose-Hubbard model in the absence of fermions, 
we here have to distinguish two cases, where a boson (bosonic hole) is 
added to a site with a fermion. Up to normalization, we have
\begin{eqnarray}
	\bigl|\psi_{+,{\cal F}}\bigr\rangle^j = {\hat a}_j^\dagger \bigl|\psi_0\bigr\rangle,\,\,\,\,
	\bigl|\psi_{-,{\cal F}}\bigr\rangle^j = {\hat a}_j \bigl|\psi_0\bigr\rangle,\,\,\,\, j\in{\cal F}, 
\end{eqnarray}
or without a fermion
\begin{eqnarray}
	\bigl|\psi_{+,{\cal N}}\bigr\rangle^j = {\hat a}_j^\dagger \bigl|\psi_0\bigr\rangle,\,\,\,\,
	\bigl|\psi_{-,{\cal N}}\bigr\rangle^j = 
	{\hat a}_j \bigl|\psi_0\bigr\rangle,\,\,\,\, j\in{\cal N}.
\end{eqnarray}
All of these vectors are eigenvectors 
of the BFH-Hamiltonian for $J_B=0$ with respective 
energies
\begin{eqnarray}
	E_{+,{\cal F}} &=& E_0 +V + U(n_0-s), \\
	E_{-,{\cal F}} &=& E_0 -V + U(n_0-s-1),\\
	E_{+,{\cal N}} &=& E_0 + U n_0, \\
	E_{-,{\cal N}} &=& E_0 + U(n_0-1), 
\end{eqnarray}
where $E_0=N \epsilon_0$.
The corresponding chemical potentials read
\begin{eqnarray}
	\mu_{+,{\cal F}}^0 &=& E_{+,{\cal F}}-E_0 = V + U(n_0-s),\\
	\mu_{-,{\cal F}}^0 &=& E_0 - E_{-,{\cal F}} = \mu_{+,{\cal F}} - U,
\end{eqnarray}
and 
\begin{eqnarray}
	\mu_{+,{\cal N}}^0 &=& E_{+,{\cal N}}-E_0 = U n_0,\\
	\mu_{-,{\cal N}}^0 &=& E_0 - E_{-,{\cal N}} = \mu_{+,{\cal N}} - U.
\end{eqnarray}
Except from the special case $V= U s$, 
the energies $E_{\pm, {\cal F}}$ and $E_{\pm ,{\cal N}}$ 
all differ from each other.  
Thus we can determine the phase boundaries
for $J_B\ne 0$ by degenerate perturbation theory within the subspaces 
$j\in {\cal F}$ and $j \in {\cal N}$ separately.

\begin{figure}[t]
\epsfig{file=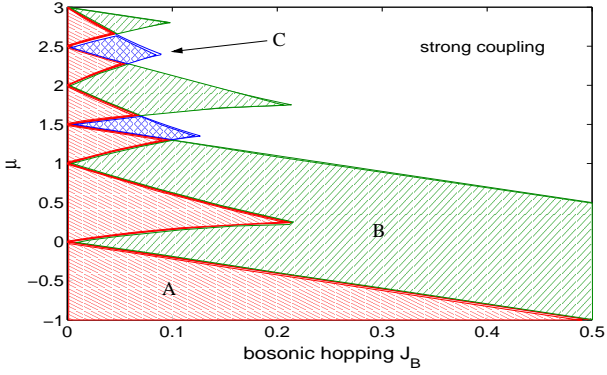,width=9cm}
\caption{Color online) Phase diagram from strong-coupling expansion and $U=1, V=1.5$.
Red areas (A) indicate truly incompressible Mott regions with gaped particle-hole excitations everywhere.
Green (B) or blue (C) areas are partially compressible quasi-Mott regions with gaped particle-hole excitation
for sites with (without) a fermion but ungapped excitation in the complementary region.}
\label{fig:StrongCouplingDiagramm}
\end{figure}

There will be a second order contribution in $J_B$ for sites $j$ that have at least one neighboring site of the same
type. For isolated sites degenerate perturbation theory will lead only to higher order terms in $O(J_B^2)$. Since the
boundaries of the incompressible phases are determined by the overall lowest-energy particle-hole excitations, we
can construct the expected phase diagram in the case of extended connected regions of fermion sites 
coexisting with extended connected regions of non-fermion sites. In this case we can directly apply the results of
Ref.\ \cite{lit:Freericks-PRB-1996} to sites with and without fermions
\begin{equation}
	\mu_{\pm,{\cal F/N}} = \mu_{\pm,{\cal F/N}}^0 + \delta
	\mu_\pm(n_0,J_B)\label{eq:strong-coupling}
\end{equation}
where
\begin{eqnarray}
	\delta\mu_+(n_0,J_B) &=& - 2 J_B(n_0+1) +J_B^2 n_0^2 \nonumber\\
	& +& J_B^3 n_0(n_0+1)(n_0+2),\\
	\delta\mu_-(n_0,J_B) &=& 2 J_B n_0  - J_B^2 (n_0+1)^2 \nonumber\\
	& -& J_B^3n_0(n_0^2-1).
\end{eqnarray}
This gives rise to two overlapping sequences of quasi-Mott lobes shifted by the boson-fermion
interaction $V$ as shown in Fig.\ \ref{fig:StrongCouplingDiagramm}.

The system is truly incompressible only in the overlap region of the quasi-Mott lobes (A). 
Points which are within one of the two sequences of quasi-Mott lobes but not in both (cases B or C) are
partially incompressible with an energy gap for a bosonic particle-hole excitation on a site
with (B) (without (C)) a fermion but without a gap for a corresponding excitation on a complementary site.
The properties of these partially incompressible phase will be discussed later.

\begin{figure}[thb]
\epsfig{file=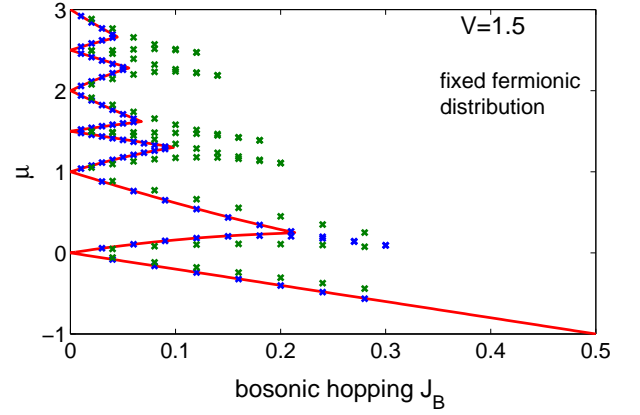,width=9cm}
\epsfig{file=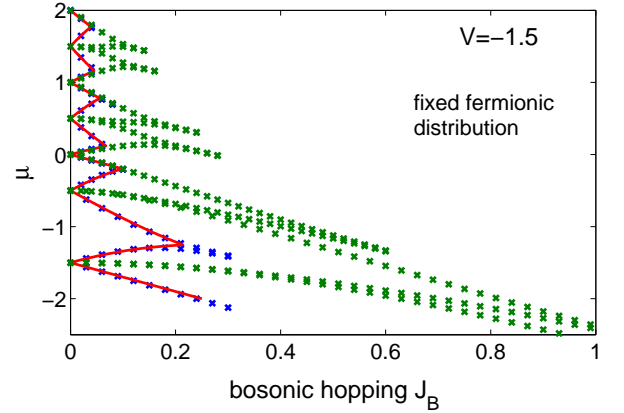,width=9cm}
\caption{(Color online) Comparison of  strong-coupling approximation (full line {\color{red} --}) and DMRG for boundaries of incompressible phases for fixed distribution of fermions corresponding to clustering (inner crosses, {\color {blue}$\pmb\times$}) or 
anti-clustering with maximum distance (outer crosses ,{\color[rgb]{0.23,0.45,0.03}$\pmb\times$}) .
 $V=1.5$ (top figure) and  $V=-1.5$ (bottom figure). $\rho_F=0.25$, and $U=1$.}
\label{fig:phases-clustering}
\end{figure}

These strong coupling results will now be complemented by 
numerical calculations using a DMRG simulation
for a system with fixed fermion distribution an open
boundary conditions. The local Hilbert space for the bosonic sector is
$\text{span}\{|0\rangle,\dots, |6\rangle\}$, so it is truncated at $6$ bosons.
The DMRG computation is done for both clustered and anti-clustered fermion distributions. The
corresponding graphs for the phase boundaries are shown in Fig.\ \ref{fig:phases-clustering}. 

\begin{figure}[thb]
\epsfig{file=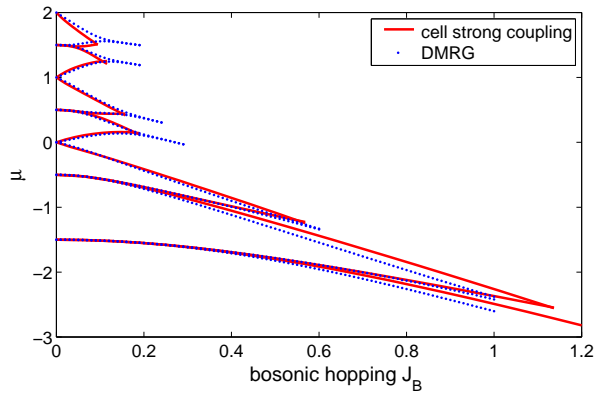,width=9cm}
\caption{(Color online) Comparison of cell-strong coupling approximation
 (full line {\color{red} --}) obtained from 
\cite{lit:Buonsante-PRA-2005} and DMRG for boundaries of incompressible phases for fixed distribution of
 fermions corresponding to anti-clustering with maximum distance (full dots,{\color{blue}$\pmb\bullet$}) .
 $V=-1.5$, $\rho_F=0.25$, and $U=1$.}
\label{fig:cell-strong-coupling}
\end{figure}

One recognizes nearly perfect agreement between numerics and strong-coupling
prediction in the case of clustering. 
This is expected since in the clustered case the majority of sites has neighbors of the same type. 
In the case
of anti-clustering, however,  the incompressible
lobes extend much further into the region of large boson hopping with a critical $J_B$ of about 
1 for a fermion filling of $\rho_F=1/4$ at $V=-1.5$. The latter 
is to be expected since in this case 
hopping to nearest neighbors is suppressed if the neighboring sites are of a different 
type ({$\cal F$} or {$\cal NF$}). 
Here the curves
of the critical chemical potential $\mu_{\rm crit}(J_B)$ that correspond to a bosonic
particle-hole excitation at a fermion site (here  $\mu_{\rm crit}(0)=-1.5, -0.5, 0.5, 1.5$ etc.)
start with a power $J_B^{\gamma}$ determined by the minimum number of hops required to reach the
next fermion site, i.e.  $\gamma=1/\varrho_F$, if $\rho_F\le 1/2$.
If the fermion filling is larger than $1/2$ the picture changes and the non-fermion
sites (hole sites) cause $\mu_{\rm crit}(J_B) \sim J_B^\gamma$ with $\gamma=1/(1-\rho_F)$. 
In principle it is possible to extend the strong-coupling perturbation expansion 
to any fermion distribution, which is however involved. 
Fig. \ref{fig:cell-strong-coupling} shows the prediction of a cell-strong coupling expansion
\cite{lit:Buonsante-PRA-2005} for an anti-clustered, fixed fermion distribution which is equivalent to
bosons in a super-lattice potential
\footnote{It should be noted that the
loop-hole insulator phases predicted for a super-lattice are for the present parameters
too small to be visible in the DMRG simulation and are expected to disappear after averaging
over disorder distributions.}.

We now want to argue that the strong-coupling expansion for a clustered fermion distribution 
provides an accurate
prediction for the boundaries of the incompressible phases in the case of {\it quenched,
random fermion disorder}. Since in the thermodynamic limit any local distribution of fermions is
realized at some places in the lattice, the actual phase boundaries 
are determined by the fermion configuration that leads to the smallest incompressible
regions. Since this is the case for a clustered fermion configuration, which in turn
is well described by the strong-coupling expansion, the latter gives a rather accurate
description of the phase transition points between compressible and incompressible
phases. 

\begin{figure}[t]
\epsfig{file=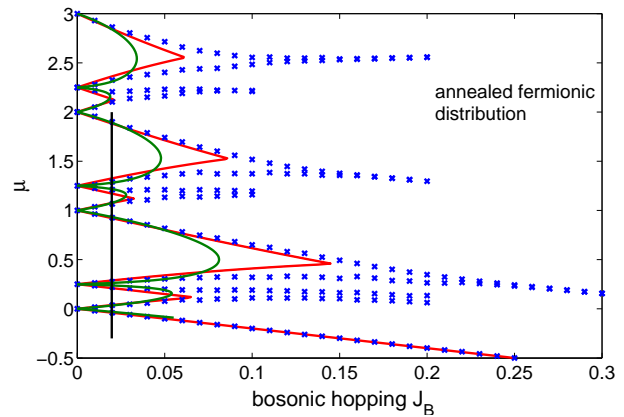,width=9cm}
\caption{(Color online) Incompressible phases for an annealed fermion distribution and  
$U=1$, $V=0.25$, $\rho_F=0.25$. Shown are the strong-coupling results (outer full line {\color{red}--}), the mean-field results from \cite{lit:Lewenstein-OptComm-2004} (inner full line {\color[rgb]{0.23,0.45,0.03}--})  and results from a DMRG calculation with $J_F=0$ (crosses {\color{blue}$\pmb \times$}). 
Vertical line indicates position of the density cut shown in Fig.\ref{fig:DensityCut} in the following section.
}
\label{fig:Phases_compare_MFT}
\end{figure}

For the case of {\it annealed} fermion distribution the strong-coupling expansion is
expected to give only less accurate results. This can be seen from Fig.\ref{fig:Phases_compare_MFT}
where we compare the predictions of the strong-coupling approximation with those from 
a DMRG simulation for annealed fermionic disorder and a mean-field ansatz.
Within the mean-field approach, e.g. of  \cite{lit:Lewenstein-OptComm-2004}, hopping is included to the system as a perturbation to the ground state
\begin{equation}
\ket{g} = \sqrt{1-\rho_F}\ket{{n}_0,0} + \sqrt{\rho_F}\ket{{n}_0-s,1}.\label{eq:groundlewenstein}
\end{equation}
Using this ground state and introducing a global bosonic order parameter $\psi$, the phase 
boundaries can be found using the usual Landau argumentation. For details see \cite{lit:Lewenstein-OptComm-2004,lit:Cramer-PRL-2004}.
Fig.\ref{fig:Phases_compare_MFT} shows the resulting phase diagram 
compared to DMRG data for annealed disorder and strong-coupling predictions.
 When comparing the different data sets one recognizes that the mean-field predictions are qualitatively correct but 
as expected only moderately precise quantitatively. It should be mentioned that the accuracy of the mean-field
approach  becomes worse even for $J_B\to 0$  for a disorder  with maximum anti-clustering.
The numerical data were obtained by letting the DMRG code freely evolve in the manifold of fermionic distributions.
The obtained distribution then gives a state which is at least close to the ground state. 
Since this procedure is prone to get stuck in local minima we checked the consistency of 
our results by implementing different sweep algorithms. In these algorithms the fermionic hopping
was not taken to be zero but was given a finite initial value which was decreased during the DMRG sweeps
to the final value zero. To ensure proper convergence we compared the data for a few representative points
($J_B=0.07$ boundaries of $(n_0=1,s=1)$ lobe; $J_B=0.15$ boundaries of $ (n_0=1,s=0)$ lobe; 
$J_B=0.03$ boundaries of $(n_0=2,s=0)$ lobe)
 to the data obtained from two different sweep strategies 
\footnote{ The sweep strategy was implemented by 
first applying an infinite size algorithm up to the system length, then applying $5$ finite size sweeps, all at 
$J_F=J_B/2$. Subsequently the hopping was reduced after a complete sweep and again $3$ sweeps 
were carried out to ensure convergence again with the new hopping amplitude. Repeatedly, the hopping 
was slightly reduced until after $30$ sweeps the fermionic hopping is set to be 0 with another $3$ sweeps.  
In the first method the hopping was reduced according to an exponential decay followed by a linear decay to zero. 
In the second method the hopping was reduced according to a cosine followed by a linear decay to zero.}. 
The difference in the chemical potential is of the order of $3\%$ independent of the sweep strategy an therefore negligible on the scale of the plot. 

\subsection{Influence of finite fermionic hopping}

\begin{figure}[htb]
\psfrag{JB}{\small$J_B$}
\psfrag{muB}{\small$\mu$}
\epsfig{file=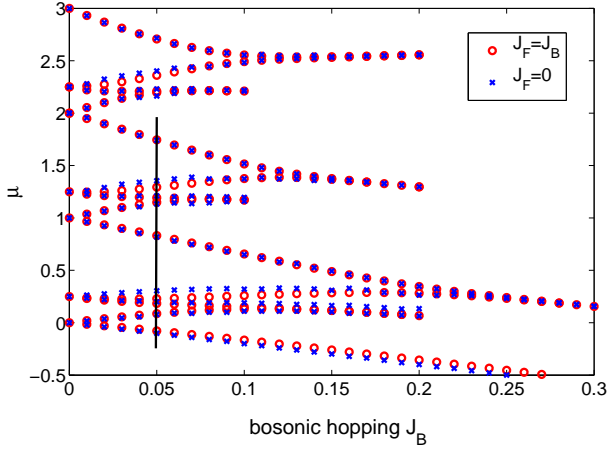,width=9cm}
\caption{(Color online) Incompressible regions for $U=1$, $V=0.25$, $\rho_F=0.25$. Shown are the results from a DMRG calculation with  $J_F=0$
and annealed fermionic disorder  ({\color{blue}$\pmb\times$}) and $J_F=J_B$ ({\color{red}$\circ$} ). The phases are the same as in fig. \ref{fig:Phases_compare_MFT}. Vertical line indicates position of the density cut shown in Fig.\ref{fig:Density2}.}
\label{fig:Phases_beideJ}
\end{figure}

The question arises how the phase diagram changes if a finite but small fermionic hopping
is included. The case $J_F\ne 0$ should be  compared to the case
$J_F=0$ for annealed fermionic disorder. Fig.\ \ref{fig:Phases_beideJ} shows a comparison
of DMRG data for $J_F=0$ and $J_F=J_B$. One recognizes that the influence of a small
fermionic hopping is rather small.

\subsection{Finite size extrapolation}

\begin{figure}[t]
\psfrag{b}{\small$J_c$}
\psfrag{a}{\small$L$}
\epsfig{file=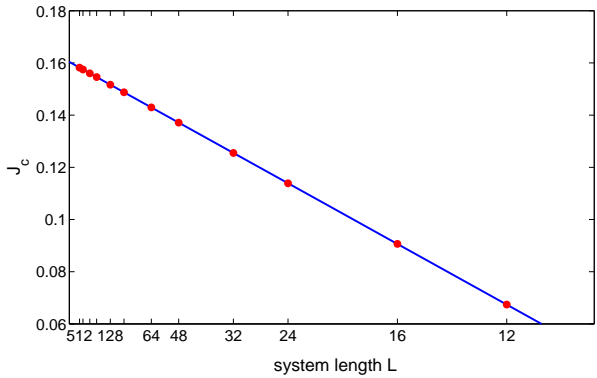,width=9cm}
\caption{(Color online) Thermodynamic limit extrapolation for the critical point 
of the $n_0=1, s=1$ lobe ($\rho_F=1/4, \rho_B=3/4$) in figure \ref{fig:Phases_beideJ}. The critical point 
is found at $J_c = 0.16038$}\label{fig:CriticalPoint}
\end{figure}

The DMRG simulations are done for finite lattices and thus finite-size
effects influence the results. To eliminate these effects each data point is obtained 
by a finite size extrapolation. This is particularly important if one wants to determine the
critical values of $J_B$ for the compressible-incompressible transition. Figure \ref{fig:CriticalPoint} 
shows the extrapolation of the tip of the lowest Mott phase in Fig.\ \ref{fig:Phases_beideJ} 
for $J_F=J_B$ to infinite lattice sizes $N\to\infty$. From a fit of $J_c$ to $\log(N)$ we find 
the critical point in the thermodynamic limit  $J_c=0.16038$. The data for different system 
lengths show the expected $1/N$ behavior shown in Ref.\ \cite{lit:Kuehner-PRB-1998} for the BHM.

\section{Partially incompressible phases}{\label{sec:BoseGlass}


\subsection{Limit of vanishing fermionic hopping}


Within the strong-coupling approximation discussed in the previous section we have
identified regions in the $\mu-J_B$ phase diagram where bosonic particle-hole excitations
are gapless if they occur on a fermion (non-fermion) site but have a finite gap on a
complementary i.e. a non-fermion (fermion) site. Associated with this is a partial 
incompressibility
\begin{eqnarray}
\frac{\partial\bigl\langle \sum_{i\in{\cal F}}{\hat n}_i\bigr\rangle}{\partial\mu}=0 ,&&\qquad 
\frac{\partial\bigl\langle {\sum_{i\in{\cal NF}}\hat n}_i\bigr\rangle}{\partial\mu}\ne 0, \nonumber\\
&&\qquad\textrm{or}\\
\frac{\partial\bigl\langle \sum_{i\in{\cal NF}}{\hat n}_i\bigr\rangle}{\partial\mu}=0, &&\qquad 
\frac{\partial\bigl\langle \sum_{i\in{\cal F}}{\hat n}_i\bigr\rangle}{\partial\mu}\ne 0. \nonumber
\end{eqnarray}
This is illustrated in Fig.\ref{fig:DensityCut}. Here the average boson  number
per site obtained from a DMRG simulation with annealed disorder 
is shown as a function of the chemical potential for constant bosonic hopping.
The curve corresponds to the parameters of Fig.\ref{fig:Phases_compare_MFT} for the vertical
cut shown in that figure at $J_B=0.02$. Also shown are
the corresponding values only for fermion sites and non-fermion sites respectively. In the partially
compressible phases the average boson number increases only for one type of sites while it stays
constant for the other. In the DMRG code the energy per particle is calculated as a function of the total
number of bosons $N$ which then yields the chemical potentials $\mu_+(N)=E(N+1)-E(N)$ and $\mu_-(N)=
E(N)-E(N-1)$.  Averaging over few values of $N$ in the compressible phase is needed here since the 
ground state fermion distribution changes with changing boson number leading to a non-monotonous dependence of $\mu$ on the boson number.

\begin{figure}[t]
\epsfig{file=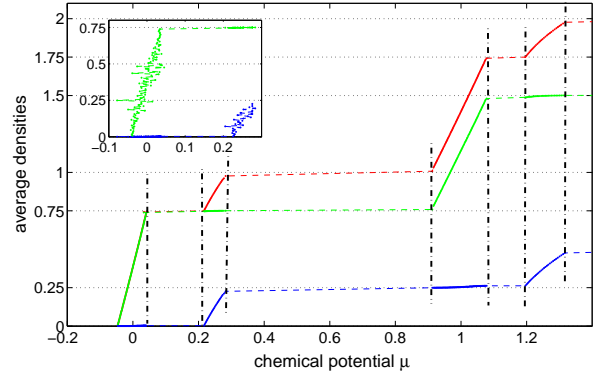,width=9cm}
\caption{(Color online) Density cut along the vertical line in figure \ref{fig:Phases_compare_MFT}. From top to bottom: overall average density ({\color{red} --}), average density for sites without a fermion ({\color[rgb]{0.23,0.45,0.03}--}), average density for sites with a fermion ({\color{blue}--}). Inset: 
Dependence of particle number at one particular site as function of chemical potential without averaging.}
\label{fig:DensityCut}
\end{figure}

We now discuss the properties of the single-particle density matrix $\langle{\hat a}_i^\dagger
{\hat a}_{i+m}\rangle$ in the partially incompressible phases. For very large values of $J_B$ the system is expected to have
a Luttinger-liquid behavior in 1d and to possess long-range off-diagonal order in higher dimensions. 
In 1d we expect that the Luttinger-liquid behavior disappears in the partially incompressible phases and that correlations decay 
exponentially. This is because in this case a single (static) impurity is 
sufficient to prevent the build-up of long-range correlations.
In higher dimensions there will be a critical fermion (or hole) filling fraction above which 
off-diagonal order is suppressed. This critical fraction is determined by percolation thresholds
and for annealed fermionic disorder depends on the actual fermion distribution in the ground state (e.g. clustered or 
anti-clustered). For a random 
fermion distribution in 2d the threshold is $\rho_F^{\rm crit}= 0.5927$ (or $1-\rho_F^{\rm crit}=0.5927$
if non-fermion sites are incompressible). The corresponding number for 3d is $\rho_F^{crit}=0.3116$.

Fig.\ref{fig:Correlations-1} shows the first-order correlations  
$\langle{\hat a}_i^\dagger {\hat a}_{i+m}\rangle$ 
as function of the distance $m$ for an {\it annealed fermion distribution} obtained from DMRG simulations
for a rather large lattice of 512 sites with incommensurate boson filling ($N_B=448$) and $\rho_F=1/4$.
For $J_B=0.07$ strong exponential decay with correlation length $l_c=1.7$ is found corresponding to
a glass-type behavior, while for $J_B=0.2$ correlations decay algebraically with $m^{-0.33}$, which
corresponds to a Luttinger liquid. Note that for the chosen boson number, which corresponds to
a non-commensurate value of $\hat Q$ there is no
incompressible phase. 

\begin{figure}[t]
\epsfig{file=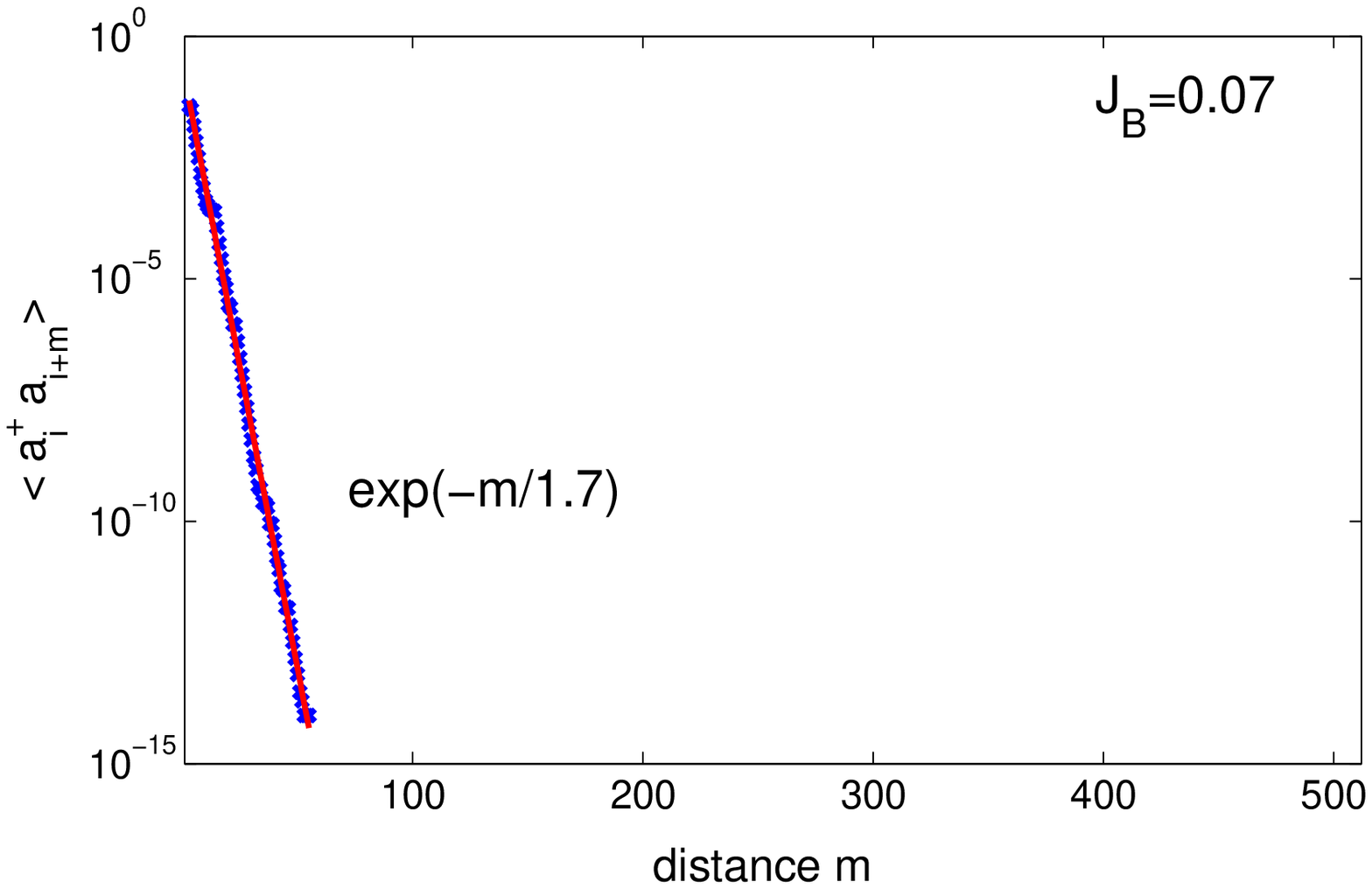,width=9cm}
\epsfig{file=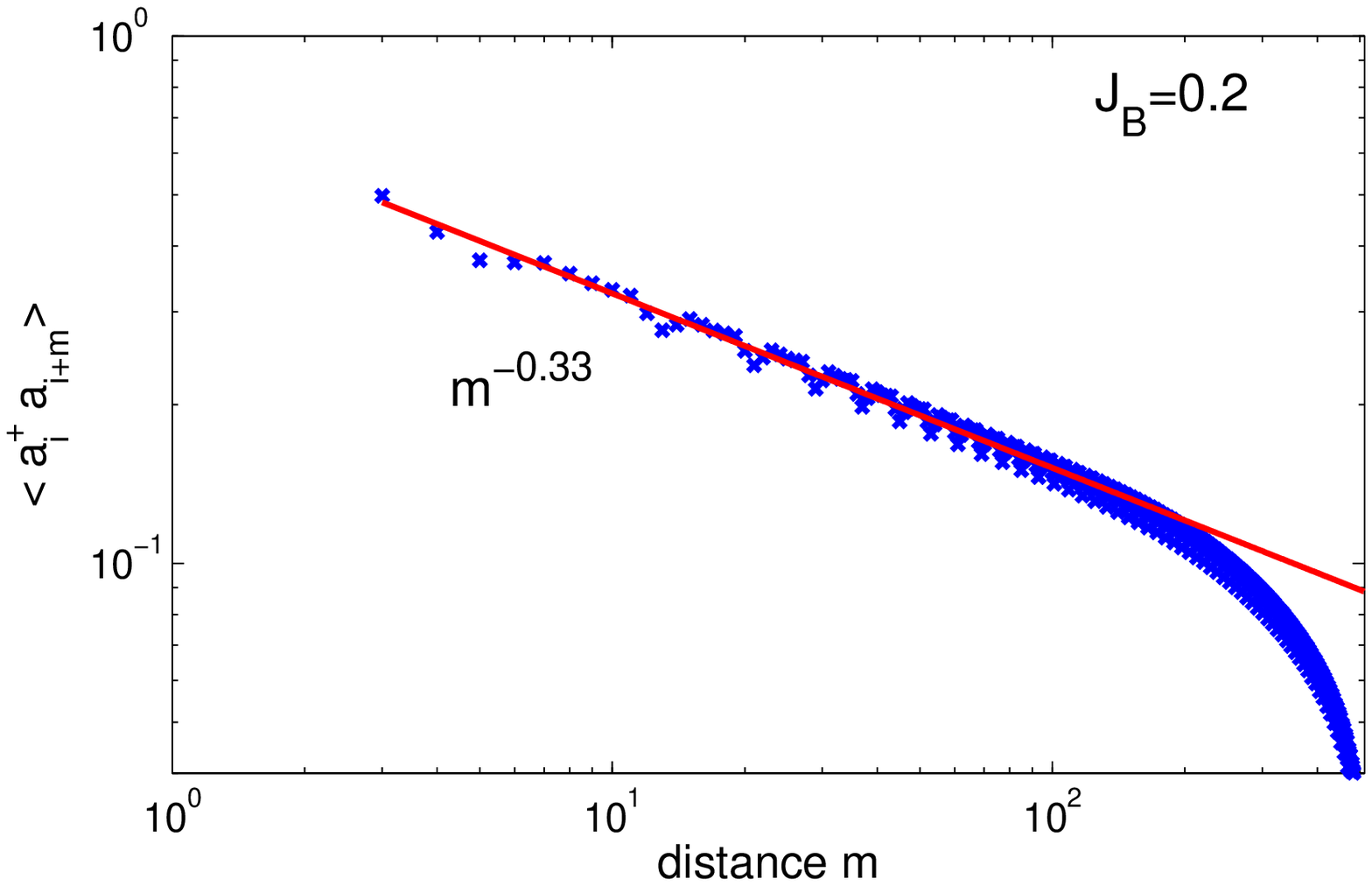,width=9 cm}
\caption{(Color online) DMRG simulations ({\color{blue}$\pmb\times$}) of first-order correlations $\langle{\hat a}_i^\dagger{\hat a}_{i+m}\rangle$ for
$V=0.25$, $\rho_F=0.25$ and $U=1$ for a lattice of 512 sites and $N_B=448$ bosons in the case of annealed disorder. (top curve:) $J_B=0.07$, line corresponds to exponential fit $\propto \exp\{-m/l_c\}$ with $l_c= 1.7$. The exponential decay for small $J_B$ is apparent. (bottom curve:) $J_B=0.2$, line corresponds to algebraic fit $\propto m^{-\alpha}$ with exponent $\alpha= 0.33$.}
\label{fig:Correlations-1}
\end{figure}

Fig. \ref{fig:Correlations-2} shows the first-order correlations 
for a {\it random, quenched fermion distribution} averaged over 100 
realizations with non-commensurate boson number ($\rho_B=N_B/N=184/128$).
Despite the sampling noise one recognizes the transition between exponential decay 
with correlation length $l_c=2.9$ for $J_B=0.03$,
 and a power-law decay with $m^{-0.37}$ for $J_B=0.2$ corresponding to a Luttinger liquid.
$J_B=0.03$ is within a partially incompressible phase, $J_B=0.2$ outside.

\begin{figure}[t]
\epsfig{file=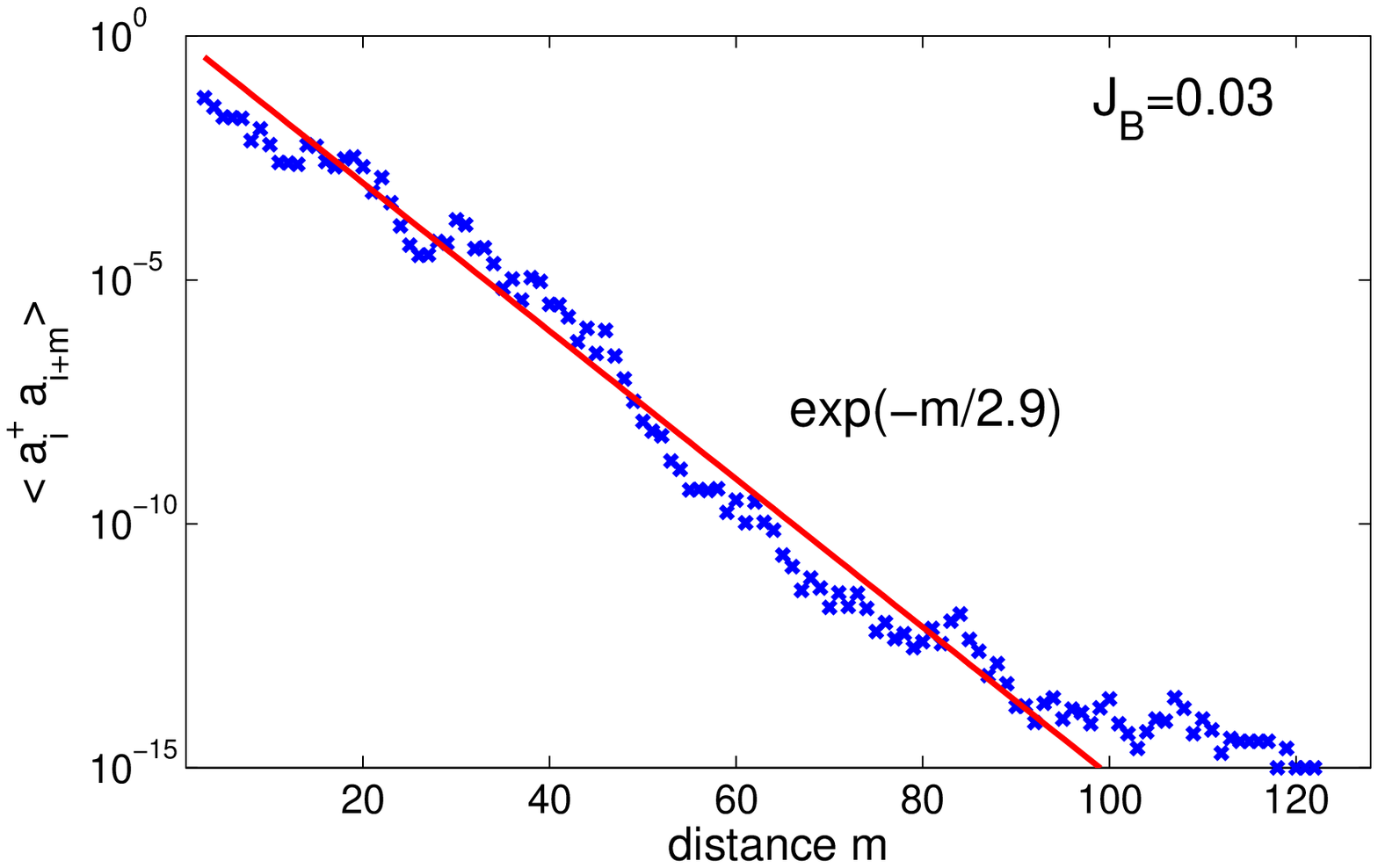,width=9 cm}
\epsfig{file=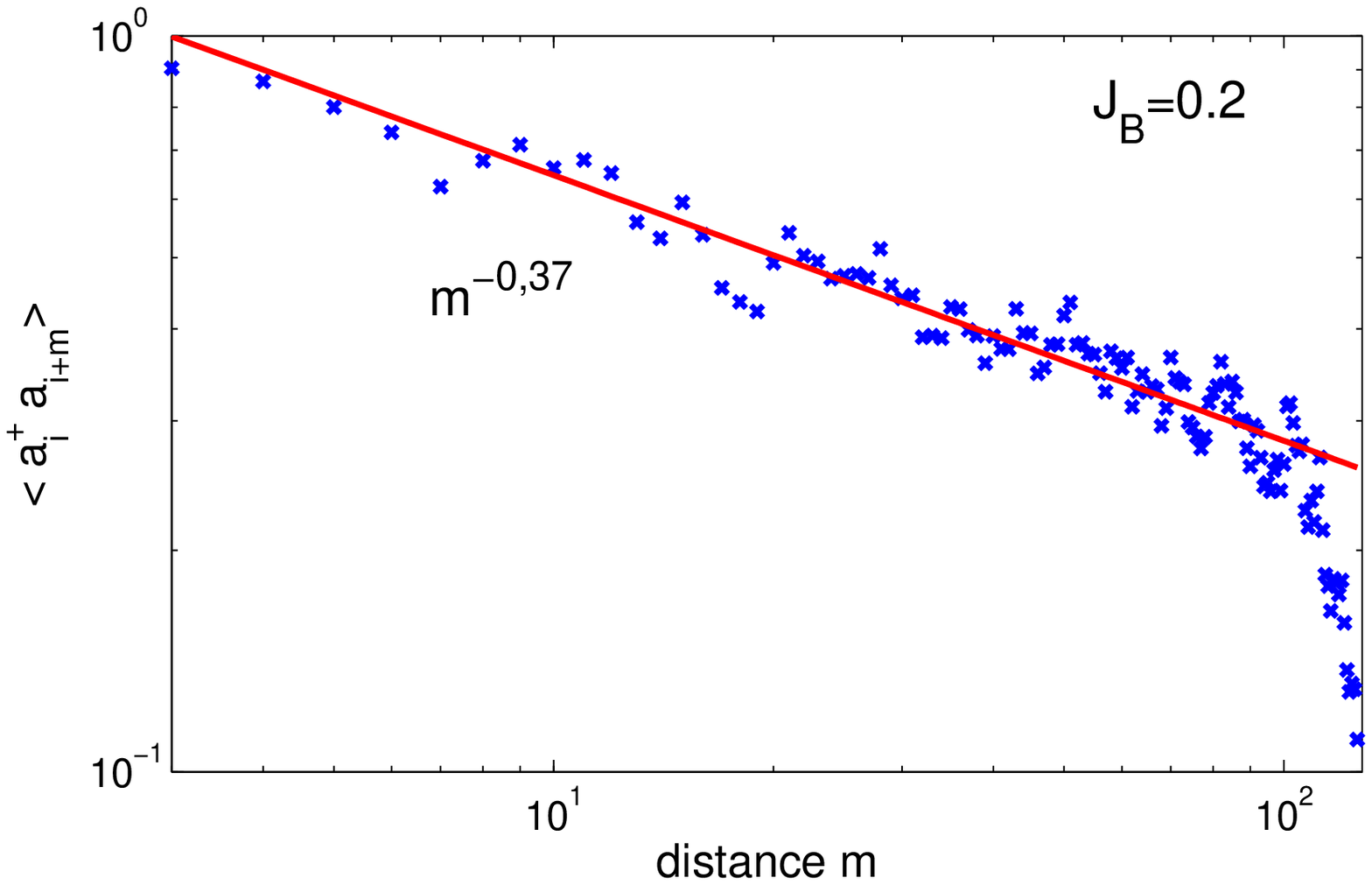,width=9 cm}
\caption{(Color online) DMRG simulations ({\color{blue}$\pmb\times$}) of first-order correlations $\langle{\hat a}_i^\dagger{\hat a}_{i+m}\rangle$ for
$V=1.5$, $\rho_F=0.375$ and $U=1$ for a lattice of 128 sites and $N_B=184$ bosons averaged over 100 fermion distributions. 
(top curve:) $J_B=0.03$, line corresponds to exponential fit
$\propto \exp\{-m/l_c\}$ with $l_c= 2.9$. The exponential decay for small $J_B$ is apparent. (bottom curve:) $J_B=0.2$, line corresponds to algebraic fit $\propto m^{-\alpha}$ with exponent $\alpha= 0.37$. To avoid finite size effects at short and long ranges, only the sites in between $13$ and $110$ are taken into account. }
\label{fig:Correlations-2}
\end{figure}

The numerical results and the above discussion indicate that the partially incompressible phases have a glass-type
character. A detailed discussion of the Bose-glass to superfluid transition will be given elsewhere \cite{lit:inprepBG}.
 

\subsection{Small fermionic hopping}

If there is a non-vanishing but small fermionic hopping, partial incompressibility is lost. Still the
increase of the boson number with increasing chemical potential at one type of sites is 
substantially less that that on the complementary type of sites.
\begin{eqnarray}
\frac{\partial\bigl\langle \sum_{i\in{\cal F}}{\hat n}_i\bigr\rangle}{\partial\mu} &\ll &
\frac{\partial\bigl\langle {\sum_{i\in{\cal NF}}\hat n}_i\bigr\rangle}{\partial\mu}\nonumber\\
&\textrm{or}&\\
\frac{\partial\bigl\langle \sum_{i\in{\cal NF}}{\hat n}_i\bigr\rangle}{\partial\mu}  &\ll &
\frac{\partial\bigl\langle \sum_{i\in{\cal F}}{\hat n}_i\bigr\rangle}{\partial\mu}.\nonumber
\end{eqnarray}
Fig.\ref{fig:Density2} shows the density cut obtained from DMRG simulations 
for the parameters of Fig.\ref{fig:DensityCut} but for $J_B=J_F=0.05$. 
It should be noted that in contrast to Fig.\ref{fig:DensityCut} averaging over sites
is not needed due to the finite mobility of the fermions.
The simulations show that the glass-type character of the 
phases survives. We expect a crossover
from glass-type to Luttinger liquid behavior with increasing
fermionic hopping. In addition due to the stronger back-action of the boson
distribution to the fermion distribution other phases such as density waves
emerge \cite{lit:Pazy-PRA-2005}. A discussion of the Bose-Fermi Hubbard model in the limit
of large fermion mobility will be given elsewhere \cite{lit:Mering-2}.

\begin{figure}[t]
\epsfig{file=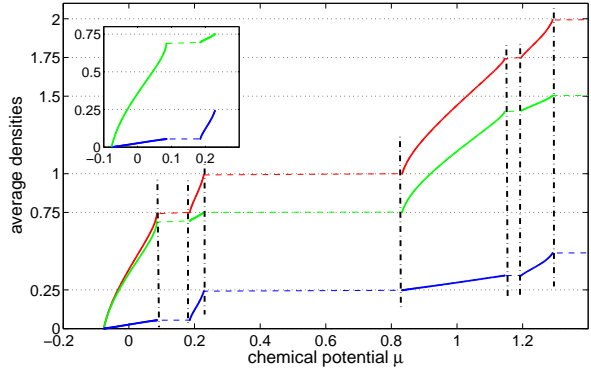,width=9cm}
\caption{(Color online) Density cut for the same parameter as Fig.\ref{fig:DensityCut} but for $J_B=J_F=0.05$ (see Fig.\ref{fig:Phases_beideJ}).
}
\label{fig:Density2}
\end{figure}


\section{fluctuations}\label{fluctuations}


In this section, we will determine the fluctuations of the bosonic number operator for vanishing fermionic hopping $J_F=0$ inside the quasi Mott lobes for quenched disorder. 
To this end, second order perturbation theory will be applied and compared to numerical results from DMRG.


For vanishing bosonic hopping the ground state with a fixed number of fermions is clearly
highly degenerate, from the distribution of $N_F$  fermions in a lattice with $N$ sites. 
For the case of quenched disorder with fixed positions of 
fermions, considered here, this degeneracy is inconsequential.
This allows to develop a tractable approach based on non-degenerate perturbation 
theory for a given fermion distribution and subsequent averaging over all of 
these distributions. In order to evaluate the fluctuations of the bosonic number operator, we hence
have to determine
\begin{equation}
	\bar n= {\mathbbm E}(\hat n_j),
\end{equation}	
which is independent of the lattice site  $j$ due to translational invariance. Here, the 
classical average
$ {\mathbbm E}$ is taken with respect to the fermionic distributions, so the average over 
the $\binom{N}{N_F}$ different distributions with equal weight.

\begin{figure}[t]
\centering
\psfrag{a}{1)}
\psfrag{b}{2)}
\psfrag{c}{3)}
\psfrag{d}{4)}
\epsfig{file=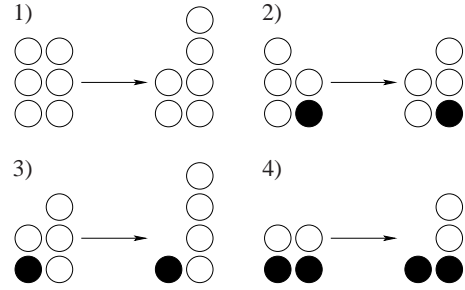,width=6cm}
\caption{Possible single-hop excitations; the energies are given by $\Delta^1E$ to $\Delta^4E$ (see text for definitions). Filled circles are fermions, open circles bosons.}\label{fig:Prozesse}
\end{figure}

We can hence proceed as in Ref.\ \cite{lit:Plimak-PRA-2004}  
to compute the fluctuations in the
boson number, for each fermion distribution, followed by the appropriate average. In second order
perturbation theory in $J_B$ at $J_F=0$, 
only bosonic hoppings to nearest neighbors contribute. 
On such two sites, clearly, four different situations can arise, dependent on whether or not
a fermion is present at each of the two sites, see Fig.\ \ref{fig:Prozesse}. The change in energy
due to these excitations is given by (here, we have no longer taken $U=1$)
\begin{eqnarray}
\Delta^{(1)}E&=&\Delta^{(4)}E=-U, \\
\Delta^{(2)}E&=&-U(1-s)-V, \\
\Delta^{(3)}E&=&-U(1+s)+V,
\end{eqnarray}
where the superscript denotes the type 
of process according to Fig.\ \ref{fig:Prozesse}. With this 
we are now able to calculate the fluctuations
	${\mathbbm E} \langle\Delta\nb^2_j \rangle$
of the bosonic number operator. 
After a number of steps, following the procedure of Ref.\ \cite{lit:Plimak-PRA-2004}, 
we find 
\begin{eqnarray}
	{\mathbbm E}  \langle\Delta\nb^2_j\rangle&=&
	2z\left(\frac{J_B}{U}\right)^2{n}_0({n}_0+1)(1-\rho_F)^2\nonumber\\
	&+& 2z\left(\frac{J_B}{U}\right)^2({n}_0-s)({n}_0-s+1)\rho_F^2\nonumber\\
	&+& 2zJ_B^2\frac{{n}_0({n}_0-s+1)+({n}_0-s)({n}_0+1)}{U^2-(Us-V)^2}\nonumber\\
	&\times & \rho_F(1-\rho_F),
	\label{eq:flukts}
\end{eqnarray}
where $z$ gives the number of nearest neighbors. 
The fluctuations show the expected quadratic dependence on the hopping strength.
Moreover, in the two limiting cases $\rho_F=0$ and $\rho_F=1$ this expression coincides with the pure 
BHM result from Ref.\
\cite{lit:Plimak-PRA-2004}. Fig.\ 
\ref{fig:Flukt} shows the analytical 
result compared with DMRG calculations
for annealed disorder. 
For small $J_B$ the agreement is rather good with 
increasing disagreement for bigger $J_B$, where 
second order perturbation theory starts to fail.

Fig.\ \ref{fig:density} shows the dependence of the fluctuations 
of the fermionic density $\varrho_F$ at a fixed hopping $J_B$. 
Also shown is one numerical curve obtained with annealed
disorder. 
The agreement between the analytical expression (\ref{eq:flukts}) 
and the numerical data shows, that the above derivation gives a 
good estimate for the fluctuations in the system for small bosonic 
hopping.

\begin{figure}[t]
\epsfig{file=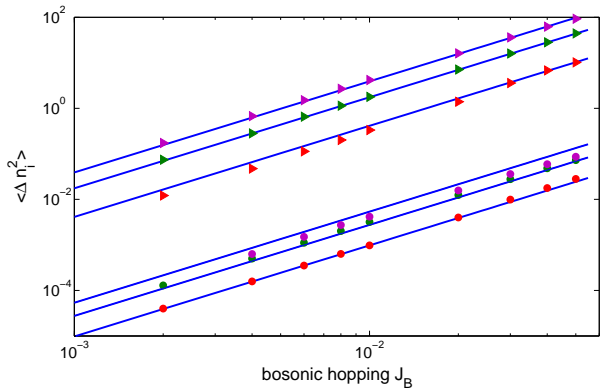,width=9cm}
\caption{(Color online) Bosonic number fluctuations in the Mott-insulating lobes for different lobes at fixed density
for annealed disorder. The upper three lines are for $V=1.25$, $\rho_F=0.375$, $s=1$ with ${n}_0 = 1$({\color{red}$\blacktriangleright$}), $2$({\color[rgb]{0.23,0.45,0.03}$\blacktriangleright$}) and $3$({\color{magenta}$\blacktriangleright$}) scaled by $10^3$. Lower three lines: $V=1.7$, $\rho_F=7/16$, $s=2$ with ${n}_0 = 2$({\color{red}$\bullet$}), $3$({\color[rgb]{0.23,0.45,0.03}$\bullet$}) and $4$({\color{magenta}$\bullet$}); solid lines are the corresponding analytic curves}
\label{fig:Flukt}
\end{figure}

\begin{figure}[t]
\centering
\psfrag{JB}{$\rho_F$}
\epsfig{file=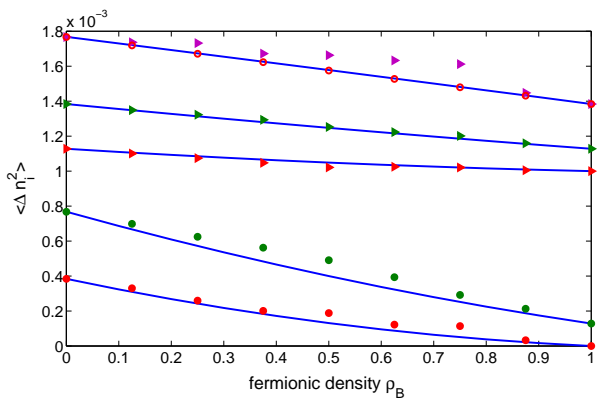,width=9cm}
\caption{(Color online) Bosonic number fluctuations in the quasi Mott lobes for different lobes at fixed hopping. The upper three lines are the results for $V=1.25$, $J_B=0.004$, $s=1$ with  ${n}_0 = 1$ ({\color{red}$\blacktriangleright$}), $2$ ({\color[rgb]{0.23,0.45,0.03}$\blacktriangleright$}) and $3$ ({\color{magenta}$\blacktriangleright$}), shifted by $10^{-3}$ upwards; lower two lines: $V=1.7$, $J_B=0.004$, $s=2$ with ${n}_0 = 2$ ({\color{red}$\bullet$}) and $3$ ({\color[rgb]{0.23,0.45,0.03}$\bullet$});  solid lines are the corresponding analytic curves. Red circles ({\color{red} $ \circ$}) are the corresponding fluctuations for a clustered disorder (only for the uppermost plot).}
\label{fig:density}
\end{figure}

\section{Summary}

In the present paper we have analyzed the $\mu - J_B$ phase diagram of the one-dimensional
semi-canonical Bose-Fermi Hubbard model with fixed number of fermions 
in the limit of vanishing fermion mobility, i.e. $J_F\to 0$. This
limit is equivalent to a Bose-Hubbard model with a random modulation of the one-site
energy. An important difference to the disordered Bose-Hubbard model \cite{lit:Fisher-PRB-1988} lies however in the
distribution of on-site energies which is here not continuous but binary, corresponding
to the presence or absence of a fermion at a given site. As a consequence there are
no extended compressible phases for vanishing bosonic hopping. Instead 
incompressible phases with in general incommensurate boson number emerge similar to the
case of a super-lattice \cite{lit:Roth-PRA-2003,lit:Buonsante-PRA-2004}. These Mott-insulating phases
which can be characterized by two integer parameter $n_0$ and $s$, denoting the number of bosons
at sites without a fermion and the shift of this number due to the presence of a fermion 
have been predicted before within mean-field and Gutzwiller approaches \cite{lit:Lewenstein-PRL-2004,lit:Cramer-PRL-2004}.
Here we determined the extend of these phases using a modified strong-coupling expansion and numerical simulations
employing the density-matrix renormalization group (DMRG). We showed that the shape of the quasi-Mott lobes depends on the
actual fermion distribution. The latter is determined by the preparation technique. If the fermionic hopping is small
but sufficiently large such that the fermions have time to find the energetically lowest configuration, one has
an annealed fermionic disorder, otherwise the distribution is random and frozen. For the annealed  case we showed that
in the limit of small, but nonzero bosonic hopping, $J_B$, the fermions form either a clustered or
an anti-clustered configuration with maximum mutual distance. A partial explanation for this behavior could be found
in terms of the composite-fermion model of \cite{lit:Lewenstein-PRL-2004}. For the case of random, quenched fermion
distributions we could derive
semi-analytic predictions for the phase boundaries of the incompressible phase using a strong-coupling
approach \cite{lit:Freericks-PRB-1996} which agreed very well with numerical simulations. Within this approach we also
identified partially compressible phases where particle-hole excitations at one type of site, i.e. either
with or without a fermion, are gap-less, while the corresponding excitations at the complementary type of sites
are gapped. The partial compressibility of these phases was verified by numerical simulations. We also showed that
that the presence of partial compressibility lead to Bose-glass phases, which are gap-less but for which first-order
correlations decay exponentially. We discussed the influence of
a finite bosonic hopping on local properties in the quasi-Mott phases using a perturbative approach supplemented by
numerical DMRG simulations. 
Finally we also discussed the influence of a finite fermionic hopping. The numerical simulations
indicate that many predictions remain valid for finite values of $J_F$ even as large as $J_B$. A more detailed
discussion of the limit of large fermionic hopping and the associated new phenomena such as density waves
etc. will be given elsewhere \cite{lit:Mering-2}.

\section{Acknowledgements*}
We would like to thank M.\ Cramer, J.\ Eisert, 
L.\ Plimak, U.\ Schollw\"ock, and M.\ Wilkens  for stimulating discussions.
We are also indebted to U.\ Schollw\"ock for providing the DMRG code and 
his support in numerical questions.
Finally we would like to thank P. Buonsante and  A. Vezzani for providing the
cell-strong coupling data for Fig.\ref{fig:cell-strong-coupling}.
This work has been supported by the DFG (SPP 1116,  GRK 792) and NIC at FZ J{\"u}lich.


\begin{thebibliography}{99}

\bibitem{lit:Bloch-RMP-2007}
	I.\ Bloch, J.\ Dalibard, and W.\ Zwerger,
	arXiv:0704.3011v1.

\bibitem{lit:Jaksch-AnnPhys-2005}
	D.\ Jaksch and P.\ Zoller, Annals of Physics, {\bf 315}, 52 (2005). 
	cond-mat/0410614v1.
	 
\bibitem{lit:Greiner-Nature-2002}
	M.\ Greiner, O.\ Mandel, T.\ Esslinger, T.W.\ H\"ansch, and I.\ Bloch, 
	Nature {\bf 415}, 39 (2002).

	
\bibitem{lit:Fisher-PRB-1988}
	M.\ P.\ A.\ Fisher, P.\ B.\ Weichman, G.\ Grinstein, and D.\ Fisher, 
	Phys.\ Rev.\ B {\bf 40}, 546 (1988).

\bibitem{lit:Jaksch-PRL-1998}
	D.\ Jaksch, C.\ Bruder, J.\ I.\ Cirac, C.\ W.\ Gardiner, and P.\ Zoller, 
	Phys.\ Rev.\ Lett., {\bf 81}, 3108 (1998).

\bibitem{lit:Lugan-PRL-2007} 
 	P.\ Lugan, D.\ Clement, P.\ Bouyer, A.\ Aspect, M.\ Lewenstein, and 
	L.\ Sanchez-Palencia,
	Phys.\ Rev.\ Lett.\ {\bf 98},   170403 (2007).

\bibitem{lit:Lewenstein-condmat-2006}
	 M.\ Lewenstein, A.\ Sanpera, V.\  Ahufinger, B.\ Damski, A.\ Sen De, and U.\ Sen,
	 cond-mat/0606771.
	 
\bibitem{lit:Scholl-EPL-1999}
	S.\ Rapsch, U.\ Schollw{\"o}ck, and W.\ Zwerger, 
	Europhys.\ Lett.\ {\bf 46}, 559 (1999).
	
	

\bibitem{lit:Inguscio-PRL-2004}
	F.\ Ferlaino, E.\ de Mirandes, G.\ Roati, G.\ Modugno, and 
	M.\ Inguscio, Phys.\ Rev.\ Lett.\ {\bf 92}, 140405 (2004).

\bibitem{lit:Esslinger-PRL-2006}
	K.\ G{\"u}nter, T.\ St{\"o}ferle, H.\ Moritz, M.\ K{\"o}hl, and T.\ Esslinger,
	Phys.\ Rev.\ Lett.\ {\bf 96}, 180402 (2006).
	
\bibitem{lit:Ospelkaus-PRL-2006}
	C.\ Ospelkaus, S.\ Ospelkaus, K.\ Sengstock, and K.\ Bongs,
	Phys.\ Rev.\ Lett.\ {\bf 96}, 020401 (2006).



\bibitem{lit:Ketterle-PRL-2002}
	Z.\ Hadzibabic, C.\ A.\ Stan, K.\ Dieckmann, S.\ Gupta, M.W.\ Zwierlein, 
	A.\ G\"oerlitz, and W.\ Ketterle, Phys.\ Rev.\ Lett.\ {\bf 88}, 160401 (2002).
	
\bibitem{lit:Albus-PRA-2003}
	A.\ Albus, F.\ Illuminati, and J.\ Eisert,
	Phys.\ Rev.\ A {\bf 68},   023606 (2003).

\bibitem{lit:Lewenstein-PRL-2004}
	M.\ Lewenstein, L.\ Santos, M.\ A.\ Baranov, 
	and H.\ Fehrmann, Phys.\ Rev.\ Lett.\ {\bf 92}, 050401 (2004).

\bibitem{lit:Lewenstein-OptComm-2004}
	H.\ Fehrmann, M.\ A.\ Baranov, B.\ Damski, M.\ Lewenstein, 
	and L.\ Santos, Opt.\ Comm.\ {\bf 243}, 23 (2004).

\bibitem{lit:Cramer-PRL-2004}
	M.\ Cramer, J.\ Eisert, and F.\ Illuminati, 
	Phys.\ Rev.\ Lett.\ {\bf 93}, 190405 (2004).

\bibitem{lit:Roth-PRA-2004}
	R.\ Roth and K.\ Burnett, 
	Phys.\ Rev.\ A {\bf 69}, 021601(R) (2004).
	
\bibitem{lit:Pazy-PRA-2005}
	E.\ Pazy and A.\ Vardi, Phys.\ Rev.\ A, {\bf 72}, 033609 (2005).
	
\bibitem{lit:Mathey-PRL-2004}
	L.\ Mathey, D.-W.\ Wang, W.\ Hofstetter, 
	M.\ D.\ Lukin and E.\ Demler, 
	Phys.\ Rev.\ Lett.\ {\bf 93}, 120404 (2004).

\bibitem{lit:Demler-PRA-2006}
	A.\ Imambekov and E.\ Demler, 
	Phys.\ Rev.\ A, {\bf 73}, 021602 (2006).

\bibitem{lit:Buechler-PRL-2003}
   	H.\ P.\ B{\"u}chler and G.\ Blatter, 
	Phys.\ Rev.\ Lett {\bf 91}, 130404 (2003).
     

\bibitem{lit:Buechler-PRA-2004}
	H.\ P.\ B{\"u}chler and G.\ Blatter, 
	Phys.\ Rev.\ A {\bf 69}, 063603 (2004).

\bibitem{lit:Pollet-PRL-2006}
	L.\ Pollet, {\it et al.},
	Phys.\ Rev.\ Lett.\ {\bf 96}, 190402 (2006.

\bibitem{lit:Pollet-condmat-2006}
	L.\ Pollet, C.\  Kollath, U.\ Schollw{\"o}ck, and M.\ Troyer, 
	cond-mat/0609604.	
	
\bibitem{lit:Freericks-PRB-1996} 
	J.\ K.\ Freericks and H.\ Monien, 
	Phys.\ Rev.\ B {\bf 53}, 2691 (1996).

\bibitem{lit:Kuehner-PRB-1998}
	T.\ D.\ K{\"u}hner and H.\ Monien, Phys.\ Rev.\ B {\bf 58}, 14741 (1998).

\bibitem{lit:Schollwoeck-RMP-2005}
	U.\ Schollw\"ock, Rev.\ Mod.\ Phys.\ {\bf 77}, 000259 (2005).

\bibitem{lit:Roth-PRA-2003}
	R.\ Roth, and K.\ Burnett, Phys.\ Rev.\ A, {\bf 68}, 023604 (2003).


\bibitem{lit:Buonsante-PRA-2004}
	P.\ Buonsante and A.\ Vezzani, Phys.\ Rev.\ A {\bf 70}, 061603(R) (2004).


\bibitem{lit:Buonsante-PRA-2005} 
	P.\ Buonsante and A.\ Vezzani, Phys.\ Rev.\ A {\bf 72}, 013614 (2005).


\bibitem{lit:QMLandau}
	L.\ D.\ Landau and E.\ M.\ Lifschitz, {\it Quantum mechanics} (Akademie-Verlag, Berlin).

\bibitem{lit:Roberts-PRL-2003}
	D.\ C.\ Roberts and K.\ Burnett, 
	Phys.\ Rev.\ Lett.\ {\bf 90}, 150401 (2003).

\bibitem{lit:inprepBG}
A. Mering, M. Fleischhauer, M. Cramer, J. Eisert and U. Schollw\"ock, (in preparation)

\bibitem{lit:Plimak-PRA-2004}
	L.\ I.\ Plimak, M.\ K.\ Olsen, and M.\ Fleischhauer, 
	Phys.\ Rev.\ A {\bf 70}, 013611 (2004).

\bibitem{lit:Mering-2}
	A.\ Mering and M.\ Fleischhauer, (to be published).

\end{thebibliography}
\end{document}